# Multilevel deconstruction of the *in vivo* behavior of looped DNA-protein complexes

Leonor Saiz and Jose M. G. Vilar[*]

Integrative Biological Modeling Laboratory, Computational Biology Program, Memorial Sloan-Kettering Cancer Center, 1275 York Avenue, Box #460, New York, NY 10021, USA

[*]Correspondence: vilar@cbio.mskcc.org

**Corresponding author:**

Jose M.G. Vilar

Computational Biology Program

Memorial Sloan-Kettering Cancer Center

1275 York Avenue, Box #460

New York, NY 10021, USA

Tel: 646-888-2603

Fax: 646-422-0717

E-mail: vilar@cbio.mskcc.org

**Manuscript information:** 21 text pages, 4 Figures, and 1 Table.

## Abstract

Protein-DNA complexes with loops play a fundamental role in a wide variety of cellular processes, ranging from the regulation of DNA transcription to telomere maintenance. As ubiquitous as they are, their precise *in vivo* properties and their integration into the cellular function still remain largely unexplored. Here, we present a multilevel approach that efficiently connects in both directions molecular properties with cell physiology and use it to characterize the molecular properties of the looped DNA-*lac* repressor complex while functioning *in vivo*. The properties we uncover include the presence of two representative conformations of the complex, the stabilization of one conformation by DNA architectural proteins, and precise values of the underlying twisting elastic constants and bending free energies. Incorporation of all this molecular information into gene-regulation models reveals an unprecedented versatility of looped DNA-protein complexes at shaping the properties of gene expression.

Many fundamental cellular processes, including transcription regulation, recombination, replication, gene silencing, and telomere maintenance, rely on the formation of DNA loops and higher order looped DNA packing structures, such as chromatin looping [1-6]. In the regulation of gene expression, proteins bound far away from the genes they control can be brought to the initiation of transcription region by looping the intervening DNA. The free energy cost of this process determines how easily DNA loops can form and therefore the extent to which distal DNA sites affect each other [4]. Assessing directly the *in vivo* value of the free energy of DNA looping is remarkably difficult, not only because the properties of the components can change when studied *in vitro*, but also because the *in vivo* probing of the cell can perturb the process under study [7].

Computational and mathematical models of gene regulation provide an avenue to connect the physical properties of DNA in its *in vivo* natural environment with the resulting cellular behavior [8,9]. This type of approach was used recently to infer the *in vivo* free energies of DNA looping by the *lac* repressor as a function of the loop length [10] from measurements of enzyme production in the *lac* operon [11], which proved to be a very accurate alternative to obtain molecular properties of the macromolecular complexes *in vivo*. The results of this analysis [10] showed that the free energy for short loops oscillates with the helical periodicity of DNA, as expected, because the operators must have the right phase to bind simultaneously to the repressor [8,12] and, unexpectedly, that the free energy in a cycle behaves asymmetrically. A Fourier analysis of the oscillations indicated that this asymmetry can be characterized by a second representative oscillatory component with a period of ~5.6 bp, in addition to the component with the *in vivo* helical period (~10.9 bp). Another striking feature of the *in vivo* free energy of looping is that the amplitude of the oscillations is as small as ~2.5 kcal/mol, similar to the typical free energy of cooperative interactions between regulatory molecules [13].

Uncovering the origin of the *in vivo* properties is important for understanding DNA looping and its effects in gene regulation, especially because current theories based on semiflexible polymer models of DNA predict symmetric and, at least, twice as large oscillations [14,15]. Different contributions, such as the anisotropic flexibility of DNA, local features resulting from the DNA sequence [16], and interactions with the *lac* repressor [17] and other DNA binding proteins, might be at play. Another potential source of complexity is the number of trajectories that DNA can follow to loop [18-21]. Thus, the observed behavior could be the result of loops with several representative conformations (Figure 1). Yet, only the lowest free energy conformation is typically considered.

Here, we develop a statistical thermodynamics approach to deconstruct the observed behavior of the expression of the *lac* operon in *Escherichia coli* cells and use it to obtain the *in vivo* properties of DNA looping by the *lac* repressor at different levels of cellular organization. At the molecular level, we propose an elastic model for DNA loop formation that considers multiple structures of the DNA-protein complex and show that, at the cellular level, the *in vivo* behavior of the free energy of looping is accurately accounted for by the presence of two distinct types of DNA loops, corresponding to two main looped DNA-protein conformations, with different relative optimal free energies, phases, and interactions with key architectural proteins. We explore in detail the effects of multiple conformations on shaping the free energy of looping DNA and the consequences that resulting free energies have for gene regulation at the cell-population level.

## RESULTS

### A multi-conformation elastic DNA model

We consider that the DNA loop can be in two distinct representative conformations (Figure 1) through the free energy of looping $\Delta G_l$, which can be expressed in terms of the free energy of each of the conformations as (see Methods)

$$\Delta G_l = -RT \ln(\sum_{i=1}^{2} \sum_{n=-\infty}^{\infty} e^{-\Delta G_{i,n}/RT}), \qquad (1)$$

where the index $i$ indicates whether the loop is in the conformation labeled 1 or 2 and $RT$ ($\approx 0.6$ kcal/mol) is the gas constant, $R$, times the absolute temperature, $T$. The integer index $n$ ranges from -infinity to +infinity and accounts for the $2\pi$ degeneracy in the twisting angle. In general, a system could have $M$ representative conformations of the nucleoprotein-DNA complex and the summation in the previous expression of $i$ would extend from 1 to $M$ (see Methods for the general case).

The free energy of a particular state includes bending and twisting contributions and is given following the classic elasticity theory of DNA [14] by

$$\Delta G_{i,n} = \Delta G_{0,i} + \frac{C}{2L}\frac{4\pi^2}{hr^2}\left(L - L_{opt,i} + n \cdot hr\right)^2, \qquad (2)$$

where $L$ is the length of the loop (in bp), $L_{opt,i}$ is the optimal spacing or phase (in bp), and $\Delta G_{0,i}$ is the corresponding optimal free energy (in kcal/mol), which depends on the type ($i$) of loop formed. In principle, the term $\Delta G_{0,i}$ could also depend on $L$ because of the bending contribution [14] but the *in vivo* results [10] indicate that it is practically constant for the range of lengths analyzed. The twisting force constant (in kcal/mol bp), $C$, and the *in vivo* helical repeat (in bp), $hr$, are considered here to be the same for the two types of loops. The free energy $\Delta G_i$ of a conformation $i$ is given by the equality

$$e^{-\Delta G_i/RT} = \sum_{n=-\infty}^{\infty} e^{-\Delta G_{i,n}/RT}$$, which includes the sum over the states of a loop conformation.

## *In vivo* free energy of DNA looping: complex average behavior from simple individual contributions

The free energy of looping $\Delta G_l$ given by Equations 1 and 2 closely reproduces the broad range of observed types of behavior (Figure 2), which consist of the *in vivo* free energies of looping DNA [10] obtained from the measured repression levels (see Methods) for two wild type situations [11,22] and a mutant lacking the architectural HU protein [22]. The *in vivo* free energies display not only asymmetric oscillations with reduced amplitude but also plateaus and secondary maxima. Therefore, our model indicates that the complex behavior of DNA looping *in vivo* emerges from a combination of the simple behavior of the individual conformations rather than from the individual conformations themselves.

The values of the parameters for the best fit (continuous black thick curves) to the data inferred from the experiments (blue symbols) are gathered in Table 1. The free energy of looping for each conformation (Figure 2 in dashed red and gray for the conformation with lowest and highest optimal free energy, respectively) depends on the length of the loop as expected for an elastic rod model of DNA, displaying symmetric oscillations with the periodicity of the DNA helix and relatively high amplitudes. The magnitudes of the amplitudes, in the order of 5 kcal/mol, are in excellent agreement with recent sequence-dependent DNA elasticity calculations for different types of *lac* repressor-DNA loops [23], which lead to oscillations of ~6 kcal/mol. Another interesting feature is the lack of a sharp increase of the looping free energy for short loops, which would be expected to arise from the bending free energy contribution. The observed behavior might originate from the high flexibility of the repressor [17] in the extended conformations [23] or from the interaction of the DNA loop with architectural proteins that help bending [24].

In both wild type situations analyzed (Figure 2, WT1 and WT2), the presence of two looped conformations (one more stable than the other by 1.0 kcal/mol and with shifts in the optimal phases of 4.3 bp or -4.2 bp) is responsible for the reduced amplitude of the oscillations and the asymmetry, including secondary maxima and/or shoulders, of the free energy curves. The inferred *in vivo* data from the two experiments is in excellent agreement with the two-conformation analysis (compare experimental blue symbols and model black thick lines in Figure 2). Our results indicate that the behavior of the *in vivo* system depends strongly on the properties of the different loop conformations, especially on the optimal free energies and optimal phases (Table 1).

Note that optimal phases and free energies between the two conformations are different for different wild type experiments (WT1 and WT2 in Table 1). These differences might arise from the differences in the experimental conditions, which are significant. For instance, the repression level in the absence of DNA looping is 135 for WT1 and 2.3 for WT2. They can also be due to potentially different boundary conditions because the loop is formed between the ideal and the main operator $O_1$ in WT1 and between the ideal and the auxiliary operator $O_2$ in WT2. The main operator is both more symmetric and 10 times stronger than $O_2$.

Optimal energies and phases determine the relative contributions of the different conformations to the observed behavior and how they change with the length of the loop. Explicitly, the probabilities for each conformation to be present, $P_1$ and $P_2$, are related to each other through the expression $P_1/P_2 = e^{-(\Delta G_1 - \Delta G_2)/RT}$, which results from the general principles of statistical thermodynamics [25]. As the distance between the two operators is changed, the less stable loop can become the most stable one. In some cases, such as for those loop lengths for which both conformations have the same free energies (when red and gray curves in Figure 2 intersect each other), the two structures are equally probable and both conformations alternate in time in a single cell and occur simultaneously in a population of cells. In the other cases, when the difference is larger than *RT*, the conformation with the lowest free energy dominates over the other one.

These two conformations of the DNA-protein looped complex, whose elastic properties we have characterized in detail, could consist of two ways of binding of the repressor to DNA, such as antiparallel and parallel DNA trajectories, which for a specific repressor conformation, i.e., the typical V-shape observed in the crystalline state [26,27], would give rise in principle to four different loop geometries [19]. Similarly, they could correspond to two different conformations of the *lac* repressor; namely, the V-shaped repressor and the extended conformation proposed from electron microscopy and fluorescence resonance energy transfer experiments in solution [28,29].

**Effects of architectural proteins**

The free energy of looping DNA *in vivo* determines the cost of forming the loop in the natural environment of the cell, which includes the double-stranded DNA molecule, the proteins that tie the DNA loops, other DNA binding proteins, and the different proteins confined within the *E. coli* cell. Architectural proteins both in eukaryotes and prokaryotes play an important role in assisting the assembly of nucleoprotein complexes and contribute to the control of gene expression as well as other DNA transactions [30-32]. These proteins locally bend or kink DNA facilitating the formation of protein-DNA looped structures [33-35] and thus are expected to affect the DNA looping properties *in vivo*. In particular, the stability of different types of looped DNA-*lac* repressor conformations has been shown to be affected by binding of the catabolite activator protein [36,37]. Other bacterial architectural proteins, such as the heat unstable nucleoid protein (HU) also referred to as histone-like protein, do not have sequence specific DNA binding sites but also bend DNA.

In the *E. coli* mutant without architectural HU protein (Figure 2, ΔHU), the *in vivo* free energy of DNA looping is compatible with the presence of two loop conformations that are similarly stable (0.2 kcal/mol difference) but have different optimal phases. In this case, the phase shift (3.3 bp) also leads to reduced amplitude of the oscillations, as in the wild-type case where HU protein is normally expressed, yet the asymmetric behavior is practically lost; now the presence of two loop conformations results in almost-symmetric oscillations with smaller amplitude. Comparison between wild type and ΔHU mutant results (Table 1) indicates that architectural proteins lower the optimal free energy of one conformation, leading to subtle differences of ~1 kcal/mol between the two most stable conformations. In systems like the Gal repressosome [38], the architectural HU protein is required to form the loop, which implies strong stabilizing effects and a single dominant conformation. In the *lac* operon, in contrast, we find that both HU stabilized and non-stabilized conformations contribute to the free energy of looping (Figure 2), which is responsible for the observed asymmetric behavior.

In all three cases studied here (Figure 2 and Table 1), the results obtained for the apparent *in vivo* twisting force constants, which also include the contributions from the repressor, are in the range 48-68 kcal/mol bp. These twisting force constants are a factor 2 smaller than the canonical value [14,39] of 105 kcal/mol bp or $2.5 \times 10^{-19}$ erg cm, and are similar to those reported recently in cyclization experiments [40].

**Shaping the behavior of the two-conformation free energy of looping**

Our analysis has shown that the complex behavior of the in *vivo* free energy of looping is accurately accounted for by combination of the rather simple behavior of two representative looped conformations (Figure 2). The major differences observed between wild type and the mutant without architectural HU protein arise mainly from the way in which the two conformations are combined; namely, from the differences in the optimal free energies and optimal phases between the two conformations. To explore the potential types of behavior that can arise when two conformations are combined, we have computed the free energy of looping by taking as reference the values of the parameters obtained for wild type (Table 1, WT1), keeping the values for one conformation fixed (conformation 1 of WT1), and systematically changing the values for the other one (Figure 3).

As the optimal free energy difference between conformations increases (Figure 3A), the behavior of the free energy changes from symmetric multiwell and wide minima, as in the ΔHU mutant, through asymmetric, typical of the wild type system, to symmetric with high amplitude oscillations (curve not shown), typical of "single-conformation" systems. A similarly broad range of types of behavior is also obtained when the difference between optimal phases changes. We have considered these changes in the context of two differences between optimal free energies: ~1.0 kcal/mol, like in wild type (Figure 3B), and ~0.0 kcal/mol, like in ΔHU mutants (Figure 3C). In both cases, as the difference between the optimal phases decreases, the amplitude of the oscillations increases. In the wild type-like situation, the oscillations of the free energy are asymmetric except for precisely tuned values of the parameters. In the ΔHU mutant-like situation, the oscillations are symmetric, and for precisely tuned values of the parameters, it is even possible to obtain oscillations with a period of half the helicity of DNA (Figure 3C, blue curve). All these results together show that the experimentally observed free energies of looping, as diverse as they are, provide just three examples of an even richer number of potential types of behavior.

## Across multiple levels: from DNA looping to gene regulation and cellular physiology

The high versatility of multi-conformation protein-DNA complexes at shaping the free energy of looping DNA propagates to the cell physiology through the effects of DNA looping in gene regulation. In a similar way as we have inferred and analyzed the *in vivo* free energy of DNA looping, we can predict the effect of a given free energy of looping on gene regulation by inverting the mathematical expression that connects the free energy of looping with the repression levels for the *lac* operon (see Methods). Explicitly, given the repression level for the system with a single operator ($R_{noloop}$), the repression level for two operators with looping follows from the free energy of looping through the expression

$$R_{loop}(L) = R_{noloop} + \frac{R_{noloop} - 1}{[N]} e^{-\Delta G_l(L)/RT} , \qquad (3)$$

where $[N]$ is the concentration of repressors.

As in the DNA looping free energy (Figure 3), the precise values of the differences in the optimal free energies and optimal phases between the two conformations strongly affect the repression level (Figure 4), leading also to a large variety of types of behaviors and degrees of repression. In general, the typical asymmetry of the free energy is less marked in the repression level (Figure 4A), to the extent that it might not be obvious in the raw experimental data, as happens in the classical experiments on the repression of the *lac* operon [11]. This loss of features leads to robust repression levels with respect to changes in the optimal phase (Figure 4B) when the optimal free energies differences are similar to the wild type value (~1 kcal/mol), whereas such robustness is not present when the optimal free energies of both conformations are similar (Figure 4C). The particular shape can thus be controlled *in vivo* by the HU architectural protein to produce either robust or sensitive gene expression patterns.

## DISCUSSION

Computational and mathematical methods provide a unique avenue to connect cellular physiology with molecular properties in a living organism [9,10,41,42]. The statistical thermodynamics approach we have developed to deconstruct the observed behavior of the expression of the *lac* operon in *E. coli* cells has allowed us to obtain the *in vivo* properties of DNA looping by the *lac* repressor at different levels of biological organization.

It was previously shown that classic experimental data on the expression of the *lac* operon in cell populations led to an unexpected, rather complex, behavior of the free energy of looping DNA *in vivo*, with small-amplitude asymmetric oscillations as a function of the length of the loop [10]. Here, we have shown that this striking behavior has its molecular origin in the ability of the *lac* repressor to loop DNA *in vivo* in at least two different ways. Thus, the intricate *in vivo* behavior of the free energy of looping is the result of combining the relatively simple behavior of each of the two looped conformations. These two types of loops have different properties and interact distinctly with the HU architectural protein. Explicitly, we found that DNA loops that interact with the HU architectural protein are ~1 kcal/mol more stable than loops that do not. Our approach has also allowed us to accurately obtain the elastic properties of the protein-DNA complexes *in vivo*, including twisting force constants, which turned out to be a factor 2 smaller than the canonical value of 105 kcal/mol bp ($2.5 \times 10^{-19}$ erg cm).

Our analysis of the effects of the molecular properties in the free energy of DNA looping at the cellular level, and their propagation to gene expression at the cell-population level, shows that there is a wide range of potential types of behavior that can arise from combining single-conformation free energies of looping. The mathematical expression for the free energy of looping (Equations 1 and 2) indicates that optimal free energies and phases, which in single-conformation systems affect only quantitative details, are key determinants of the resulting behavior. In particular, the asymmetry in the oscillations is the consequence of the presence of a slightly preferred loop conformation with different optimal phase. Symmetric oscillations in the free energy result from equally stable loop conformations, a strongly dominant conformation, or conformations with the same optimal phases. In *E. coli* cells, as shown by our results, the HU protein preferentially affects one loop conformation making it slightly more stable, thus leading to the observed asymmetry.

Different loop trajectories have been observed *in vitro* for diverse nucleoprotein complexes [19,21]. In particular, *in vitro* experiments of DNA cleavage by the *Sfi*I endonuclease, a type II restriction endonuclease that binds to two DNA sites as a tetramer by looping out the intervening DNA, have shown coexistence and alternative conformations [21] as the DNA spacer between binding sites is changed for loop sizes of 109-170 bp. They also observed similar periodicities for the two conformations as well as different phases in *in vitro* electrophoresis experiments. There are also studies on the Gal repressor showing that several non-simultaneous trajectories can exist and that there is a single configuration of the complex for a particular loop length when the HU protein is present [38]. Our results provide evidence that shows, for the first time, that alternative and simultaneous nucleoprotein-DNA configurations are present *in vivo*.

At the cell-population level, whether the typical asymmetry of the free energy propagates to the repression level is controlled by the values of the optimal free energies and phases. In general, the asymmetry in the repression level is less marked than in the free energy, to the extent that it might not be obvious in the raw experimental data [11].

Our results indicate that the biological consequences of having two or more DNA-looped conformations include a reduced dependence on the positioning of the DNA binding sites. For instance, by combining two DNA conformations, it is possible to reduce the amplitude of the typical oscillations in the free energy as a function of the length of the loop from ~5 kcal/mol to ~1 kcal/mol (Figure 3C). In this way, DNA appears to the cell to be much more malleable than it actually is in a single conformation. The presence of multiple DNA conformations also provides an extra layer of control of the properties of gene regulation. In the case of the *lac* operon, we have shown that the HU protein stabilizes one DNA conformation. Similarly, it has also been shown that the Catabolite Activator Protein (CAP) stabilizes preferentially certain loop conformations [37]. Thus, expression of CAP, HU, and other architectural proteins can change the DNA looping properties in a conformation-dependent manner and select the precise details of the interactions between distal DNA sites.

In broad terms, our analysis has revealed that the formation of DNA loops *in vivo* is tightly coupled to the molecular properties of the proteins and protein complexes that form the loop. There is a high versatility of looped DNA-protein complexes at establishing different conformations in the intracellular environment and at adapting from one conformation to another. This versatility underlies the unanticipated behavior of the *in vivo* free energy of DNA looping and can be responsible not only for asymmetric oscillations with decreased amplitude but also for plateaus and secondary maxima. All these features indicate that the physical properties of DNA can actively be selected to control the cooperative binding of regulatory proteins and to achieve different cellular behaviors.

# METHODS

## *Free energy of DNA looping from multiple conformations*

Following the statistical thermodynamics approach [25], the free energy of looping, $\Delta G_l$, can be expressed in terms of the free energy for each individual conformation as

$$e^{-\Delta G_l/RT} = e^{-\Delta G_1/RT} + e^{-\Delta G_2/RT} + e^{-\Delta G_3/RT} + ...,$$

where the right hand side of the equation has as many terms as the number of possible representative conformations of the looped DNA-protein complex. In practice, only the conformations with lowest free energy will have a significant effect in the observed behavior. In particular, we have shown that typically only two distinct conformations contribute significantly, and thus $e^{-\Delta G_l/RT} = e^{-\Delta G_1/RT} + e^{-\Delta G_2/RT}$, which leads to $\Delta G_l = -RT \ln(e^{-\Delta G_1/RT} + e^{-\Delta G_2/RT})$ for the free energy of DNA looping.

## *In vivo free energy of DNA looping from physiological measurements*

The *in vivo* free energy of DNA looping by the *lac* repressor's binding to the main and an auxiliary operator can be expressed in terms of the measured repression levels through a well-established model for gene regulation by the *lac* repressor [9]. For the experimental conditions consisting of a strong auxiliary operator, which are those of the experiments considered here, the free energy of looping DNA [10] for an inter-operator distance $L$ is given by:

$$\Delta G_l(L) = -RT \ln \frac{R_{loop}(L) - R_{noloop}}{R_{noloop} - 1}[N],$$

where $R_{loop}(L)$ is the measured repression level, a dimensionless quantity used to quantify the extent of repression of a gene; $R_{noloop}$ is the repression level in the absence of DNA looping; $[N]$ is the concentration of repressors; and $RT$ is the gas constant times the absolute temperature.

## Figure and Table Legends

**Figure 1**: Two plausible alternative loop conformations of the *lac* repressor-DNA complex. The bidentate repressor, with the two dimers that form the functional tetramer shown in red, simultaneously binds DNA, colored orange, at two sites. The two structures represent two plausible trajectories of the DNA loop and two plausible conformations of the *lac* repressor (V-shaped and extended).

**Figure 2**: Two-conformation analysis of the *in vivo* free energy of DNA looping. The *in vivo* free energy of looping DNA by the *lac* repressor (blue symbols) was obtained as described in Saiz et al. [10] (see also Methods) from the measured repression levels of Muller et al. [11] for wild type (WT1) and of Becker et al. [22] for wild type (WT2) and a mutant that does not express the architectural HU protein (ΔHU). As repression levels in the absence of looping (see Methods and Saiz et al. [10]) we have used 135 (WT1), 2.3 (WT2), and 1.7 (ΔHU). The thick black continuous lines correspond in each case to the best fit to the free energy $\Delta G_l$ given by Equations 1 and 2, which considers the contributions of two looped conformations. The contributions of each conformation are shown separately as red ($\Delta G_1 = -RT \ln(\sum_{n=-\infty}^{\infty} e^{-\Delta G_{1,n}/RT})$) and black ($\Delta G_2 = -RT \ln(\sum_{n=-\infty}^{\infty} e^{-\Delta G_{2,n}/RT})$) dashed lines. The values of the parameters for the best fit are shown in Table 1.

**Figure 3**: Free energy of looping for a two-conformation elastic DNA model. Different types of behavior are obtained by changing two key parameters: the difference in optimal free energies ($\Delta G_{0,1} - \Delta G_{0,2}$) and optimal phases ($L_{opt,1} - L_{opt,2}$). (A) The difference in optimal free energies between the two configurations increases from 0 kcal/mol (blue) to 1.5 kcal/mol (red) in increments of 0.5 kcal/mol whereas the difference in optimal phases is kept fixed at 4.2 bp. (B, C) The difference in optimal phases between the two conformations increases from -5.5 bp (blue) to 0 bp (red) in increments of 5.5/3 bp whereas the difference in optimal free energies is kept fixed at 1 kcal/mol (B) and at 0 kcal/mol (C).

**Figure 4**: Effects in gene expression of the free energy of looping for a two-conformation elastic DNA model. Repression levels obtained with Equation 3 using the corresponding free energies of Figure 3. (A) Differences in optimal free energies are varied and the optimal phases are kept fixed. (B, C) Differences in optimal phases are varied and optimal free energies are kept fixed at two different values: 1 kcal/mol (B) and 0 kcal/mol (C).

**Table 1:** *In vivo* values of the molecular parameters of the looped DNA-*lac* repressor complex. The data shows the best fit values of the parameters of the model with two distinct looped DNA-*lac* repressor conformations (Equations 1 and 2) to the *in vivo* free energies obtained from Muller et al. experiments [11] for wild type (WT1) and from Becker et al. experiments [22] for wild type (WT2) and a mutant that does not express the architectural HU protein (ΔHU).

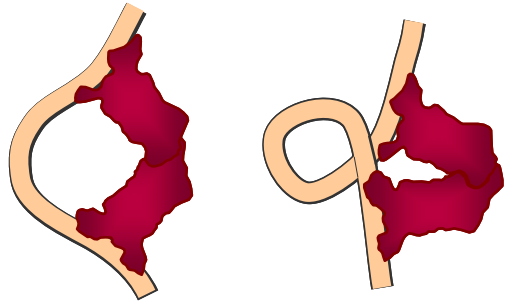

**Figure 1**

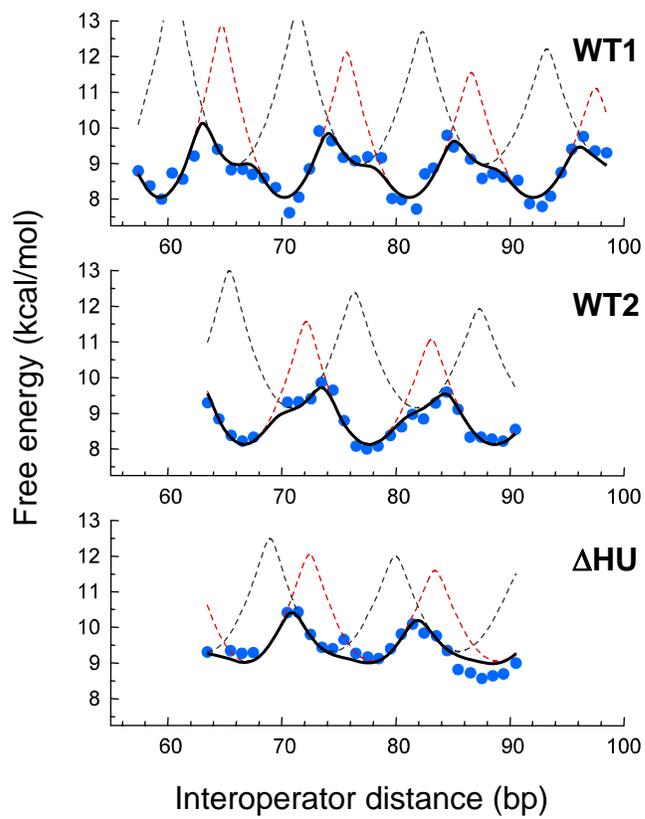

Figure 2

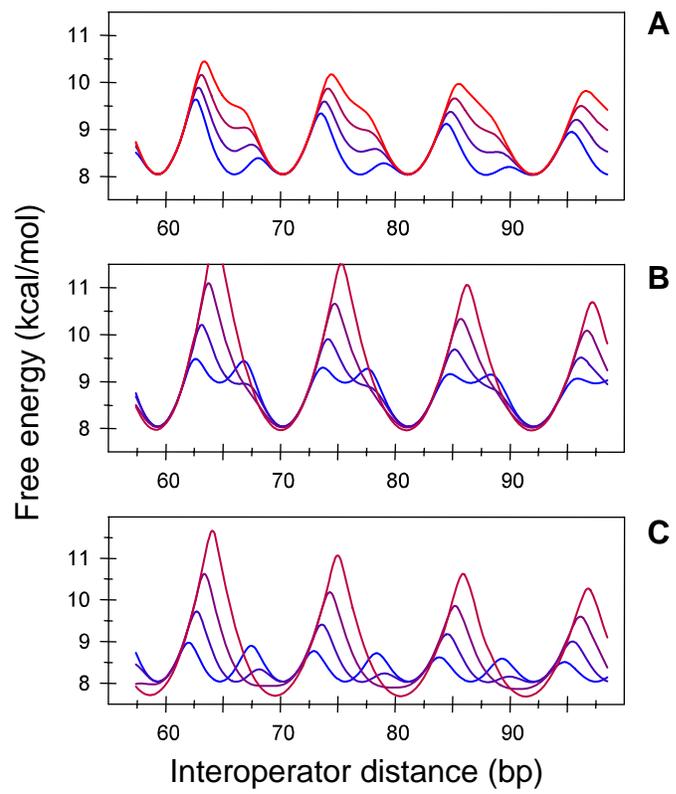

Figure 3

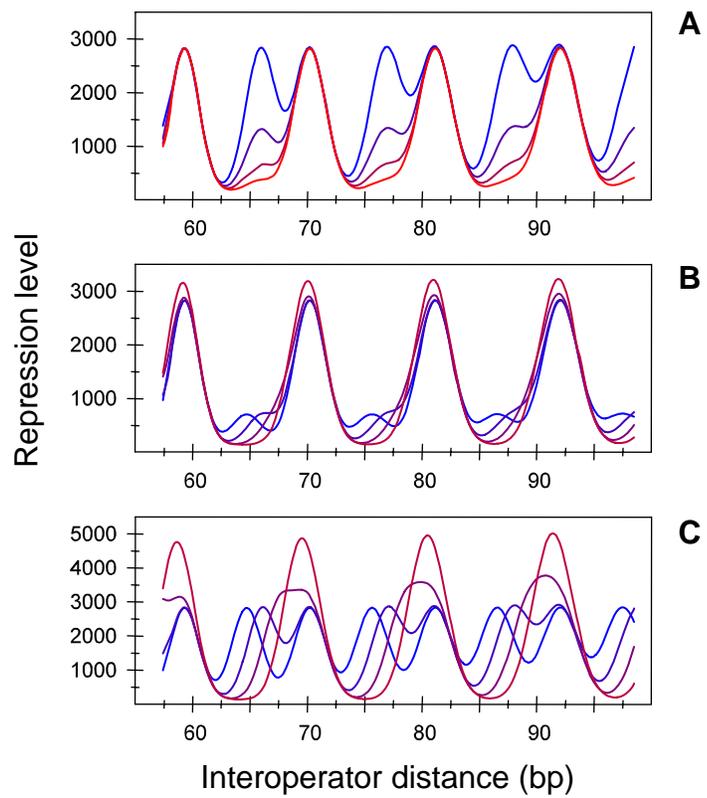

**Figure 4**

|     | $\Delta G_{0,1}$ (kcal/mol) | $\Delta G_{0,2}$ (kcal/mol) | $L_{opt,1}$ (bp) | $L_{opt,2}$ (bp) | $hr$ (bp) | $C$ (kcal/mol bp) |
|-----|------|------|-----|------|------|-----|
| **WT1** | 8.0 | 9.0 | 4.7 | 0.4 | 10.9 | 68 |
| **WT2** | 8.1 | 9.1 | 1.0 | 5.2 | 11.0 | 55 |
| **ΔHU** | 9.1 | 9.3 | 1.1 | -2.3 | 11.0 | 48 |

**Table 1**